\documentclass{jkas}

\def\year{2018} % publication year
\def\month{December} % publication month
\def\volume{51} % journal volume
\def\issue{6} % journal issue
\def\beginpage{185} % first page of article
\def\endpage{195} % last page of article
\def\received{September 30, 2018} % date paper was received by JKAS
\def\accepted{November 24, 2018} % date of acceptance
\date{Received \received; accepted \accepted}

\def\kms{~{\rm km~s^{-1}}}
\def\cm3{~{\rm cm^{-3}}}

\def\muG{{\mu\rm G}}

\usepackage{flushend}

\title{Re-Acceleration of Fossil Electrons by Shocks \\Encountering Hot Bubbles in the Outskirts of Galaxy Clusters}

\author{Hyesung Kang}

\affil{Department of Earth Sciences, Pusan National University, 2 Busandaehak-ro, Geumjeong-gu, Busan 46241, Korea \email{hskang@pusan.ac.kr}}

\def\corrauthor{H. Kang}

\def\runningauthor{Kang}

\def\runningtitle{Re-Acceleration of Fossil Electrons by ICM Shocks}

\def\keywords{acceleration of particles --- cosmic rays --- galaxies: clusters: general --- shock waves}

\def\abstracttext{
Galaxy clusters are known to host many active galaxies (AGNs) with radio jets, which could expand to form radio bubbles with relativistic electrons in the intracluster medium (ICM).
It has been suggested that fossil relativistic electrons contained in remnant bubbles from extinct radio galaxies can be re-accelerated to radio-emitting energies by merger-driven shocks via diffusive shock acceleration (DSA),
leading to the birth of radio relics detected in clusters.
In this study we assume that such bubble consist primarily of thermal gas entrained from the surrounding medium and dynamically-insignificant amounts of relativistic electrons.
We also consider several realistic models for magnetic fields in the cluster outskirts, including the ICM field that scales with the gas density as $B_{\rm ICM}\propto n_{\rm ICM}^{0.5}$.
Then we perform time-dependent DSA simulations of a spherical shock that runs into a lower-density but higher-temperature bubble with the ratio $n_{\rm b}/n_{\rm ICM} \approx T_{\rm ICM}/T_{\rm b}\approx 0.5$.
We find that inside the bubble the shock speed increases by about 20 \%, but the Mach number decreases by about 15\% in the case under consideration.
In this re-acceleration model, the observed properties of a radio relic such as radio flux, spectral index, and integrated spectrum would be governed mainly by the presence of seed relativistic electrons and the magnetic field profile as well as shock dynamics. Thus it is crucial to understand how fossil electrons are deposited by AGNs in the ICM and how the downstream magnetic field evolves behind the shock
in detailed modeling of radio relics.
}

\begin{document}
\jkashead %% set title, authors, abstract, etc.
%--------------------------------------------------------------------

\section{Introduction\label{intro}}

Radio jets from active galactic nuclei (AGNs) are observed to excavate X-ray cavities or radio bubbles on scales of $\sim 10-100$ kpc in the intracluster medium (ICM) \citep[e.g.,][]{mcnamara07,fabian2012,tadhunter16}.
Low-power, ``edge-darkened" FRI radio galaxies are more numerous than high-power, ``edge-brightened" FRII radio galaxies \citep{fanaroff74},
so FRIs are expected to play more important roles in governing the thermal evolution of the ICM.
Observed morphology of radio jets can vary from wide-angle-tailed (WAT), narrow-angle-tailed (NAT), to head-tail (HT) radio galaxies, depending mainly on the ram pressure of the ICM bulk flows impinging on the jets.

The typical structures of jets in FRI radio galaxies from the inside out consist of (1) relativistic inner jets on parsec scales, (2) flaring regions on kpc scales decelerated to sub-relativistic speeds by turbulent entrainment, and (3) outer recollimated jets on $\sim10$ kpc scales \citep[e.g.,][]{laing2002,laing2013}.
These jets often develop diffuse radio lobes or plumes extending to $\sim 100$~kpc \citep[e.g.,][]{feretti99,hardcastle2004}.
Due to entrainment of the interstellar medium of the host galaxy and the surrounding ICM, radio lobes/plumes are likely to contain significant amounts of thermal gas in addition to relativistic plasmas originating from the jets \citep[e.g.,][]{croston14}.

Radio relics are diffuse radio sources detected in the outskirts of merging clusters of galaxies \citep[see][for a review]{brunetti2014}.
They are also known as `radio shocks', since the radio emission is thought to come from relativistic electrons accelerated at ICM shocks
via Fermi first order process \citep{vanweeren10}.
Recently, the Merging Cluster Collaboration\footnote{www.mergingclustercollaboration.org} compiled a panchromatic (from radio to X-ray) atlas of merging clusters with radio relics,
which can be used to study the particle acceleration at merger-driven shocks as well as the merger geometry and dynamics \citep{golovich18}.

In the {\it test-particle} regime, the energy spectrum of cosmic-ray electrons (CRe) accelerated by collisioness shocks takes a power-law form, $N(E) \propto E^{-s}$,
where $s=2(M_{\rm s}^2+1)/( M_{\rm s}^2-1)$ and $M_{\rm s}$ is the shock sonic Mach number \citep{dru83}.
The synchrotron radiation spectrum due to these CRe then has a power-law form
$j_{\nu}(x_s)\propto \nu^{-\alpha_{\rm sh}}$, where the spectral index is $\alpha_{\rm sh}= (s-1)/2$ \citep[e.g.,][]{kang11}.
Thus a `radio Mach number', $M_{\rm rad}=[(3+2\alpha_{\rm sh})/(2\alpha_{\rm sh}-1)]^{1/2}$, can be inferred from the spectral index of these radio shocks.
On the other hand, the shock Mach number can be also estimated from the temperature discontinuity obtained from X-ray observations,
using the shock jump condition, $T_2/T_1=(M_{\rm X}^2+3)(5M_{\rm X}^2-1)/16/M_{\rm X}^2$, where the subscripts, $1$ and $2$, identify the upstream and downstream states, respectively.
%---------------------------------------------------------------
\begin{figure}[t!]
\centering
\includegraphics[trim=2mm 2mm 2mm 2mm, clip, width=80mm]{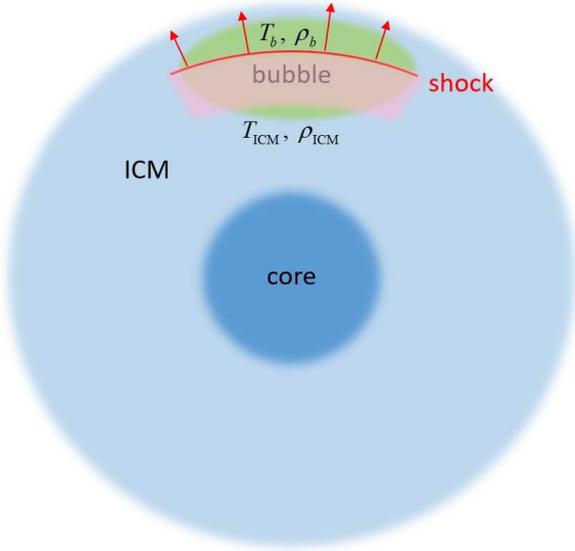}
\caption{Schematic diagram showing a shock encountering a hot bubble (green) with low-energy fossil relativistic electrons in the outskirts of a galaxy cluster.
Radio-emitting region (pink) behind the shock contains high-energy cosmic-ray electrons. \label{fig1}
}
\end{figure}
%------------------------------------------------------------

Although $M_{\rm rad}$ and $M_{\rm X}$ agree reasonably well for most observed radio shocks, in a few cases observed values of $\alpha_{\rm sh}$ and the X-ray temperature jump imply that $M_{\rm rad} > M_{\rm X}$
\citep[e.g.,][]{akamatsu15}. In the case of the Toothbrush relic, for instance, it was proposed that $M_{\rm rad}\approx 2.8$ \citep{vanweeren12},
while  $M_{\rm X}\sim 1.2-1.5$ \citep{vanweeren16}.
Using particle-in-cell (PIC) plasma simulations, \citet{guo2014} argued that electrons can be accelerated at $M_{\rm s}=3$ quasi-perpendicular ICM shocks up to the Lorentz factor $\gamma_e \lesssim 10$.
However, \citet{kangetal18} has recently shown through similar PIC simulations that weak {\it subcritical} shocks with $M_{\rm s} \lesssim 2.3$ cannot pre-accelerate electrons from the thermal pool to
high enough non-thermal energies that they can participate in the Diffusive Shock acceleration (DSA) processes. This is because the fraction of the incoming
electrons satisfying the reflection criteria for shock drift acceleration is far too small, and the self-generation of oblique waves via the firehose
instability is ineffective at such subcritical shocks. 
Further, as argued in \citet{hong15}, the $M_{\rm X}$ value for the Toothbrush relic maybe underestimated due to projection effects and complex shock 
morphology. 
So we do not consider subcritical shocks ($M_{\rm s}<2.3$) in this study.

In the so-called re-acceleration model for radio relics, `fossil' electrons may provide seed electrons to be injected to the DSA process at ICM shocks \citep[e.g.,][]{ensslin98,ensslin01,kang11,kang12,pinzke13,kang16a,kang16b}.
Relativistic electrons from dead radio jets may cool down to
$\gamma_e \lesssim 300$ with characteristic cooling time scales of $\sim 10^9$~yrs via synchrotron and inverse Compton losses.
Those low-energy CRe are not radio-luminous in typical microgauss magnetic fields and referred as `fossil electrons' \citep[e.g.,][]{kang16a}.
Radio-emitting electrons re-accelerated to $\gamma_e \gtrsim 10^4$ by merger-driven shocks may become giant Gischt relics
such as the Sausage relic in the cluster CIZA J2242.8+5301 \citep{vanweeren10} and the Toothbrush relics in the cluster 1RXS J060303.3 \citep{vanweeren12}.
The introduction of fossil CRe alleviates the following problems of the DSA model for radio relics:
(1) low DSA efficiency expected for weak shocks with $M_{\rm s}\lesssim 3$,
(2) low frequency of merging clusters with detected radio relics, compared to the expected occurrence of
ICM shocks,
and (3) shocks detected in X-ray observations without associated radio emission \citep{kangetal17}.

%---------------------------------------------------------------
\begin{figure*}[t!]
\centering
\includegraphics[trim=5mm 125mm 5mm 5mm, clip, width=150mm]{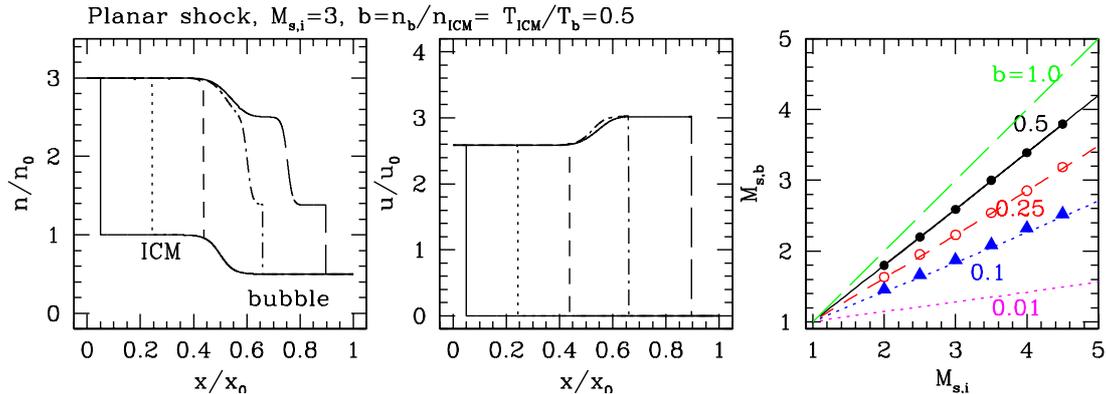}
\caption{Left and Middle panels: Time evolution of a planar shock with $M_{\rm s,i}=3.0$ running in to a hotter bubble with
$b=n_{\rm b}/n_{\rm ICM}=T_{\rm ICM}/T_{\rm b}=0.5$. Here the shock is moving to the right and the boundary between the ICM and the bubble changes gradually with a hyperbolic tangent profile.
The shock speed begins to increase at the third time epoch (dashed lines), but the shock Mach number decreases due to higher sound speed as it enters the hotter bubble.
Right panel: Shock Mach number inside the bubble, $M_{\rm s,b}$, estimated from Equation (\ref{bubeq}), versus the initial shock Mach number, $M_{\rm s,i}$
for $b=1$ (green long-dashed line), $b=0.5$ (black solid), $b=0.25$ (red dashed), $b=0.1$ (blue dotted), and $b=0.01$ (magenta dotted).
The values of $M_{\rm s,b}$ inferred from 1D hydrodynamic simulations are shown with black filled circles for $b=0.5$, red open circles for $b=0.25$, and blue filled triangles for $b=0.1$. \label{fig2}
}
\end{figure*}
%---------------------------------------------------------------

Merging clusters with giant Gischt relics and double radio relics commonly host low-power radio galaxies: for instance,
PLCKG287.0+32.9 \citep{bonafede14}, Abell 3411-3412 \citep{vanweeren17a},
MACS J0717.5+3745 \citep{vanweeren17b}, and CIZA J2242.8+5301 \citep{digennaro18}.
In this study, we assume that radio lobes and plumes from those radio galaxies, transported outward by the buoyant force, survive long after the jets from host AGNs were turned off, and become bubbles (i.e., lower density and higher temperature compared to the surrounding ICM) with relativistic electrons in the cluster outskirts.
{ Although the long-term evolution of these remnant bubbles is largely unknown, we presume for simplicity that the hot gas cools down and mixes with fossil relativistic electrons
through turbulent entrainment, while maintaining overall structural integrity as diffuse clouds.}

In \citet{kangetal17} and \citet{kang17} we attempted to reproduce the observed radio flux density profiles, $S_{\nu}$, spectral index profiles,
$\alpha_{\nu}$, and the volume-integrated spectra, $J_{\nu}$,
of the Toothbrush relic and the Sausage relic. We use a re-acceleration model in which a spherical shock of sonic Mach number,
$M_{\rm s}\approx 3$, sweeps through a finite-sized cloud with fossil electrons.
There we assumed that the cloud is composed of the ICM thermal gas and small amounts of fossil electrons (i.e., the pressure ratio of $P_{\rm CRe}/P_{\rm gas} \sim 10^{-3}-10^{-4}$), since there are no observational signatures indicating the presence of an X-ray cavity hotter than the surrounding ICM in those clusters.

In the present study, we explore a feasible scenario, in which the shock encounters a remnant cloud of an old radio bubble created by an extinct radio galaxy, in pressure equilibrium with the surrounding ICM (i.e.,
$n_{\rm b} T_{\rm b} = n_{\rm ICM} T_{\rm ICM}$) in the cluster outskirts (see Figure 1).
Hereafter, the subscript `b' is used to represent physical parameters for the bubble.
Inside the bubble, the shock sonic Mach number, $M_{\rm s,b}$, decreases due to the higher sound speed, while the X-ray surface brightness decreases due to the lower gas density.
The bubble is assumed to be filled with small amounts of `radio-dark' fossil electrons with $\gamma_e \lesssim 300$.

The magnetic field behind the shock, $B_{\rm dn}$, is another key parameter that determines the surface brightness profiles of radio shocks, because both synchrotron emissivity and cooling depend on it.
{In our previous studies, we adopted simple models in which the preshock magnetic field strength, $B_1$, is isotropic and constant, while its transverse components are compressed across the shock transition.} Moreover, the downstream magnetic field energy decreases behind the shock with the gas pressure, i.e., $B_{\rm dn} = B_2 (P/P_2)^{1/2}$,
where $P_2$ and $B_2$ are the gas pressure and magnetic field strength immediate downstream of the shock \citep[e.g.,][]{kang16a,kang16b,kangetal17}.
According to the Faraday rotation measure study by \citet{bonafede10}, magnetic field strength decreases radially as $B_{\rm ICM}\propto n_{\rm ICM}^{0.5}$ in the cluster outskirts,
whereas the ICM density typically scales as $n_{\rm ICM}\propto r^{-2.5}$ \citep[e.g.,][]{ghirardini18}.
The magnetic field inside these bubbles, $B_{\rm b}$, could be stronger than or similar to $B_{\rm ICM}$ of the surrounding ICM.
Although magnetic field is expected to be amplified mainly through compression across low $M_{\rm s}$ shocks \citep[e.g.,][]{ji2016},
the amplification or decay of magnetic fields behind weak ICM shocks by MHD  turbulence is not well understood \citep[e.g.,][]{donnert16}.
Thus we consider several magnetic field models in this study.

In the next section, the numerical simulations and the shock models are described.
The DSA simulation results are presented and discussed in Section~\ref{results},
followed by a brief summary in Section~\ref{summary}.

\section{Numerical Calculations\label{numerics}}

\subsection{Planar Shock Case\label{plane}}

In order to understand how shock dynamics changes, when it enters into a low-density (but high-temperature) bubble, we first study one dimensional (1D)
planar shocks.
\citet{pfrommer11} considered the 1D Riemann problem for a planar shock that runs into a low-density bubble with
the density parameter, $b=n_{b}/n_{\rm ICM}=T_{\rm ICM}/T_{b}<1$.
The ratio of the shock Mach number in the bubble to that in the ICM, $\mu = M_{\rm s,b}/M_{\rm s,i}$ can be found by solving Equation (B4) of that paper.
If the adiabatic index is  $\gamma_g=5/3$ for both the ICM and the bubble, the equation becomes
\begin{eqnarray}
1= {{\mu^2 M_{\rm s,i}^2-1}\over {\mu b^{1/2}(M_{\rm s,i}^2-1)}} - {{[(5M_{\rm s,i}^2-1)(M_{\rm s,i}^2+3)]^{1/2}} \over {M_{\rm s,i}^2-1}} \nonumber\\
 \times [1- ( {{5\mu^2 M_{\rm s,i}^2-1}\over {5M_{\rm s,i}^2-1}} )^{1/5}].
\label{bubeq}
\end{eqnarray}

The right panel of Figure 2 shows the solutions of Equation (\ref{bubeq}) for $b=0.01-1.0$.
Although Equation (\ref{bubeq}) is a nonlinear equation that can be solved only numerically, it turns out that
we can adopt the following approximate linear-fitting form:
\begin{equation}
M_{\rm s,b} \approx b^{-1/3} M_{\rm s,i} + (1 -b^{-1/3})
\end{equation}
for $0.1 \lesssim b \le 1$.
For $b=0.01$, on the other hand, $M_{\rm s,b} \approx 0.14 M_{\rm s,i} + 0.86$ for $M_{\rm s,i}\lesssim 5$.
In the case of $M_{\rm s,i}=4.0$, $M_{\rm s,b}\approx 2.26$, 2.85, and, 3.39 for $b=0.1$, 0.25, and 0.5, respectively.
Thus the presence of a hotter bubble results in a weaker shock, which in turn leads to a steeper radio synchrotron spectrum from the shock accelerated electrons.

\citet{croston14} examined the particle content of the jets and plumes of the radio galaxy 3C31 and suggested that on $10-100$~kpc scales
the thermal pressure of the gas entrained from surrounding medium dominates over the combined pressure of radio plasma and magnetic fields.
Their model predicted that the entrained gas in the plumes would be heated significantly to around $100$~keV.
The bubble considered here could be the cooled remnant of such a plume,
{mixed with the ICM gas through turbulent entrainment and} transported from the cluster core to the outskirts ($d\sim 1-1.5$~Mpc).
Noting that no X-ray cavities associated with observed giant radio relics have been detected, we assume that the density ratio of our bubble could
be only moderate, e.g., $b\sim 0.5 $.
For instance, the X-ray emissivity, $j_{\rm X}\propto n_{\rm b}^2 T_{\rm b}^{1/2} \propto b^{3/2}$ decreases by a factor of $\sim 3$ for $b=0.5$.
We also assume that the thermal gas dominates over fossil relativistic electrons, so $\gamma_g=5/3$ for both the ICM and the bubble.

The left and middle panels of Figure~\ref{fig2} shows the time evolution of a Mach 3 planar shock running into a lower density bubble with $b=0.5$ in 1D hydrodynamic simulations.
Here we consider a bubble with boundary that changes gradually with a hyperbolic tangent profile, so a contact discontinuity or other complex structures are not generated when the shock runs into the bubble.
As can be seen in the figure, the shock compression ratio, $\sigma = \rho_2/\rho_1$, decreases but
the shock speed increases as it enters into the bubble.

In the right panel, the shock Mach number inside the bubble, $M_{\rm s,b}$, estimated from the 1D simulations, are shown as blue filled triangles for $b=0.1$,
red open circles for $b=0.25$, and black filled circles for $b=0.5$, as a function of the initial Mach number, $M_{\rm s,i}$.
This demonstrates that $M_{\rm s,b}$ estimated from the simulations agree well with the solutions of Equation (\ref{bubeq}).

\subsection{DSA Simulations for 1D Spherical Shocks\label{DSA}}

The numerical setup for our DSA simulations was described in detail \citet{kangetal17}.
Some basic features are repeated here in order to make this paper self-contained.

We follow the electron DSA along with radiative cooling and postshock turbulent acceleration
by solving the diffusion-convection equation in the 1D spherically symmetric geometry:
\begin{eqnarray}
& &{\partial g_{\rm e}\over \partial t} + u {\partial g_{\rm e} \over \partial r} \nonumber\\
& &= {1\over{3r^2}} {{\partial (r^2 u) }\over \partial r} \left( {\partial g_{\rm e}\over
\partial y} -4g_{\rm e} \right)
+ {1 \over r^2}{\partial \over \partial r} \left[r^2 \kappa(r,p){\partial g_{\rm e} \over \partial r} \right] \nonumber\\
& &+ p {\partial \over \partial y}\left[ {D_{pp} \over p^3} \left( {\partial g_{\rm e}\over \partial y} -4g_{\rm e} \right) \right]
+ p {\partial \over {\partial y}} \left( {b\over p^2} g_{\rm e} \right),
\label{diffcon}
\end{eqnarray}
where $f_e(r,p,t)=g_e(r,p,t)p^{-4}$ is the pitch-angle-averaged phase space distribution function
for CR electrons, $u(r,t)$ is the flow velocity, $y=\ln(p/m_e c)$, $m_e$ is the electron mass, and $c$ is
the speed of light \citep{skill75}.
Here $r$ is the radial distance from the cluster center.
In addition, the gasdynamic conservation equations are solved for spherically expanding shocks as described in \citet{kj06}.

We adopt a Bohm diffusion coefficient, $\kappa(p)= \kappa_N (p/m_ec)$ for relativistic electrons,
where the normalization factor, $\kappa_N= m_ec^3/(3eB)= 1.7\times 10^{19} {\rm cm^2s^{-1}}/B_{\muG} $,
with $B_{\muG}$ expressed in units of $\muG$.
The electron energy loss term, $b(p)= \dot p_{\rm Coul} + \dot p_{\rm sync+iC} $, accounts for Coulomb scattering,
synchrotron emission, and iC scattering off the cosmic background radiation.

In merging clusters, turbulence can be injected into the ICM and cascades down to smaller scales,
which may further energize relativistic electrons via stochastic Fermi II acceleration \citep{brunetti2014}.
Similarly, MHD/plasma turbulence can be generated at collisionless shocks, leading to { turbulent dynamo of magnetic fields and turbulent acceleration of particles}
in the postshock region of radio relics.
As in \citet{kangetal17}, we consider Fermi II acceleration via the transit time damping resonance due to compressive fast modes of postshock turbulence,
which is parameterized as $D_{pp} = { p^2 /{4\ \tau_{\rm acc}}}$ with $\tau_{\rm acc}\approx 10^8$~yrs \citep{brunetti2007}.
It was demonstrated that such levels of postshock turbulent acceleration is required in order to reproduce the broad radio flux density profiles
of the Toothbrush and Sausage relics \citep{kangetal17,kang17} .

%---------------------------------------------------------------
\begin{figure*}[t!]
\centering
\includegraphics[trim=5mm 120mm 5mm 5mm, clip, width=150mm]{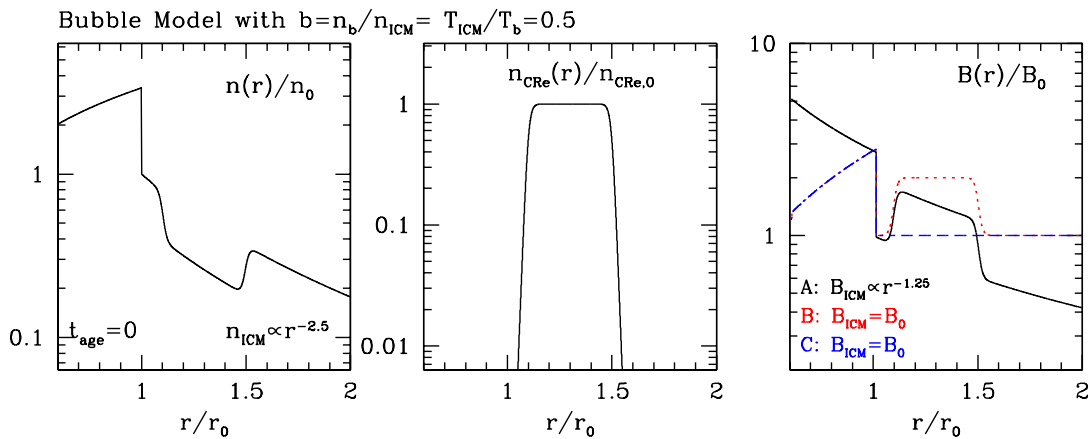}
\caption{Initial conditions for the gas density, fossil CR electron density, and magnetic field strength for the model that contains a hotter bubble with $b=0.5$. The shock is located at $r/r_0=1.0$ at $t_{\rm age}=0$. See Section \ref{Bfield} for the details of Model A (black solid line), B (red dotted line), and C (blue dashed line) for the magnetic field profiles.\label{fig3}
}
\end{figure*}
%---------------------------------------------------------------

\subsubsection{Bubble in the Cluster Outskirts\label{bubble}}

We assume that the shock encounters a cloud/bubble that contains fossil CRe with the following power-law spectrum
with exponential cutoff:
\begin{equation}
f_{\rm pre}(p) = f_o \cdot p^{-s} \exp \left[ - \left({p \over p_{e,c}} \right)^2 \right],
\label{fpre}
\end{equation}
where the slope $s=4.5$ and the cutoff $\gamma_{e,c}=300$ are adopted.
However, the exact shape of $f_{\rm pre}(p)$ is not important, because these low-energy CR electrons serve only as seed particles to be
injected to the DSA process.
The normalization factor, $f_o$, can be parameterized by the ratio of the pressure of fossil electrons
to the gas pressure in the bubble, $N \equiv P_{\rm CRe}/P_{\rm b} \sim 10^{-4}$, which is a free parameter to be adjusted to reproduce observed flux levels.

The ICM in the cluster outskirts is assumed to be described by $n_{\rm ICM}\propto r^{-2.5}$ and $T_{\rm ICM} \propto r^{-0.5}$ \citep[e.g.,][]{ghirardini18}.
Considering postshock length scales of $\sim 100-150$~kpc for giant radio relics, the shock must sweep up at least $300-450$~kpc in distance.
For example, if the shock propagates radially outward from $0.8$~Mpc to $1.2$~Mpc in the cluster halo, the preshock density and temperature decrease by factors of $2.8$ and $1.2$, respectively.

At the onset of the simulations ($t_i$), the shock with initial shock Mach number, $M_{\rm s,i}=4.0$ runs into a bubble located at $r_0=0.8$~Mpc.
The gas density and temperature profiles of the ICM are given as $n_{\rm H} = 10^{-4}\cm3 (r/r_0)^{-2.5}$ and $k T= 2.7 ~{\rm keV}(r/r_0)^{-0.5}$, respectively.
We define the ``shock age'', $t_{\rm age} \equiv t - t_i$, as the time since the beginning of the simulations.

The bubble is realized by the constant density ratio, $b=n_{\rm b}(r)/n_{\rm ICM}(r)$, with hyperbolic tangent boundaries (see Figure 3)
{ to prevent the formation of a contact discontinuity and multiple shocks.}
The temperature in the bubble is specified by the pressure equilibrium condition, $n_{\rm b} T_{\rm b}= n_{\rm ICM} T_{\rm ICM}$.
The density of fossil electrons, $n_{\rm CRe}$, is assumed to be constant inside the bubble with hyperbolic tangent boundaries, as shown in the middle panel of Figure~\ref{fig3}.

\begin{figure*}[t!]
\centering
\includegraphics[trim=3mm 70mm 6mm 10mm, clip, width=160mm]{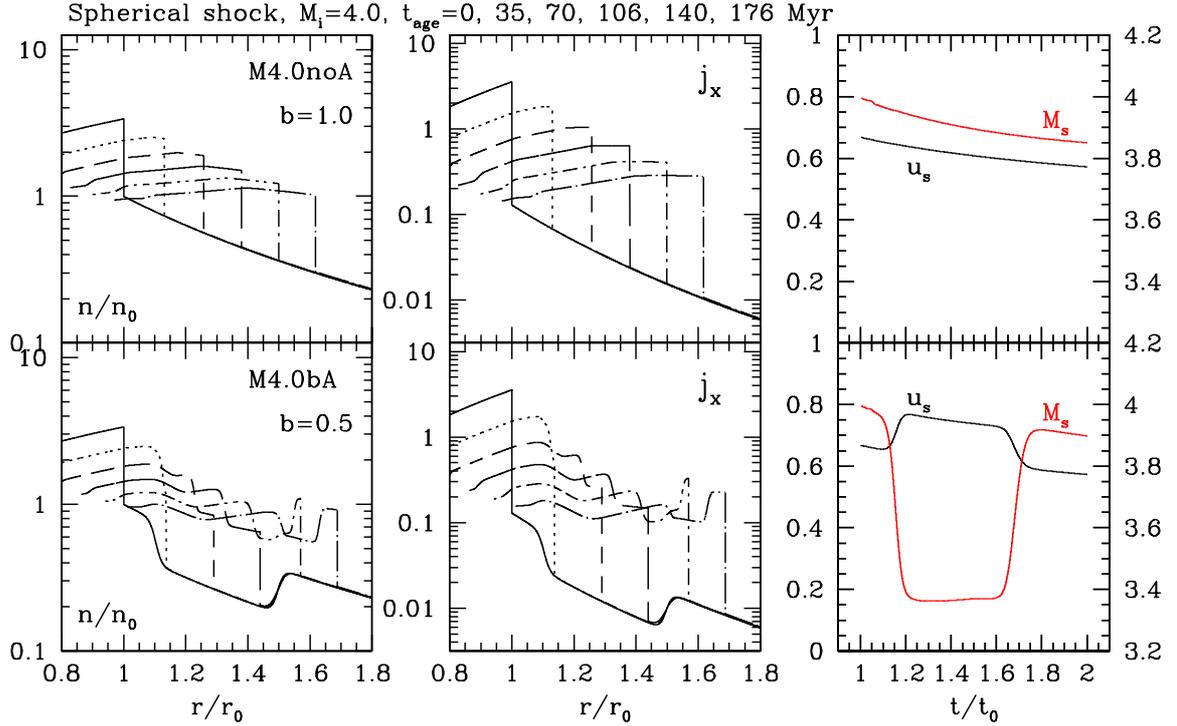}
\caption{Time evolution of $M_{\rm s,i}=4.0$ shocks without ($b=1.0$, upper panels) and with ($b=0.5$, lower panels) a bubble.
The gas density (left panels) and X-ray emissivity (middle panels) are shown at six different time epochs.
Right panels: Evolution of the shock speed, $u_s(t)/u_0$ (black lines, the left-hand axis), and Mach number, $M_{\rm s}(t)$ (red lines, the right-hand axis) in the two models.
Here, $r_0=0.8$~Mpc, $t_0=176$~Myr, and $u_0=4.4\times 10^3 \kms$.
Inside the bubble the shock speeds up but the Mach number decreases due to the higher sound speed.\label{fig4}
}
\end{figure*}

Figure~\ref{fig4} shows how the presence of a bubble with the density ratio $b=0.5$ influences shock dynamics. The shock expands and slows down slightly during the period shown in the figure.
The upper panels show the evolution of the shock coasting radially the ICM halo (M4.0noA), while the lower panels show how the shock interacts with the hotter bubble (M4.0bA). See Table~\ref{table1} for the model names and parameters.
Inside the bubble, the shock speed increases, but its Mach number, the shock compression ratio, and X-ray emissivity, $j_{X} \propto n^2 T^{1/2}$, all decrease.
As shown in the right panel of Figure~\ref{fig4},
in the M4.0noA model the shock slows down slightly and the shock Mach number decreases only to $M_{\rm s} \approx 3.9$ at $r_s/r_0 \sim 1.2-1.4$,
while it reduces by $\sim 15$~\% to $M_{\rm s,b}\approx 3.35$ inside the bubble in the M4.0bA model with $b=0.5$.
As a result, the radio spectral index at the shock position (relic edge), $\alpha_{\rm sh}$, becomes larger (or steeper) inside the bubble.

\begin{table}[t!]
\caption{Model Parameters for Spherical Shocks\label{table1}}
\centering
\begin{tabular}{lcccc}
\toprule
Model     & $M_{\rm s,i}$ & $b=n_{\rm b}/n_{\rm ICM}$ & $B(r)$ model$^{\rm a}$ & Remarks\\
\midrule
M4.0noA    & 4.0 & 1.0  &A  & no bubble\\ \addlinespace
M4.0bA    & 4.0 & 0.5  &A &  \\
M4.0bB    & 4.0 & 0.5  &B & \\
M4.0bC    & 4.0 & 0.5  &C & \\
\bottomrule
\addlinespace
\end{tabular}
\\$^{\rm a}$  See Section~\ref{Bfield} for magnetic field models.
\end{table}

\subsubsection{Magnetic Field Models\label{Bfield}}

Since we solve the gasdynamic equations instead of MHD equations, the evolution of magnetic fields has to be modeled by hand in the simulations.
Three different components of magnetic field profiles need to be specified: (1) $B_{\rm ICM}$ for the background ICM, (2) $B_{b}$ inside the bubble,
and (3) $B_{\rm dn}$ for the downstream flow behind the shock (see the right panel of Figure 3).
The ICM magnetic field strength is assumed to scale with $n_{\rm ICM}$ as $B_{\rm ICM} \propto n_{\rm ICM}^{0.5} \propto r^{-1.25}$ \citep[e.g.,][]{bonafede10}.
The field strength inside the bubble is assumed either to scale with the density of fossil electrons, $B_{b} \propto n_{\rm CRe}$,
or to scale inversely with the gas density,  $B_{b} \propto n_{\rm b}^{-1}$ (i.e., stronger magnetic fields inside the bubble).

%---------------------------------------------------------------
%---------------------------------------------------------------

%---------------------------------------------------------------
\begin{figure*}[t!]

\centering
\includegraphics[trim=3mm 70mm 6mm 10mm, clip, width=160mm]{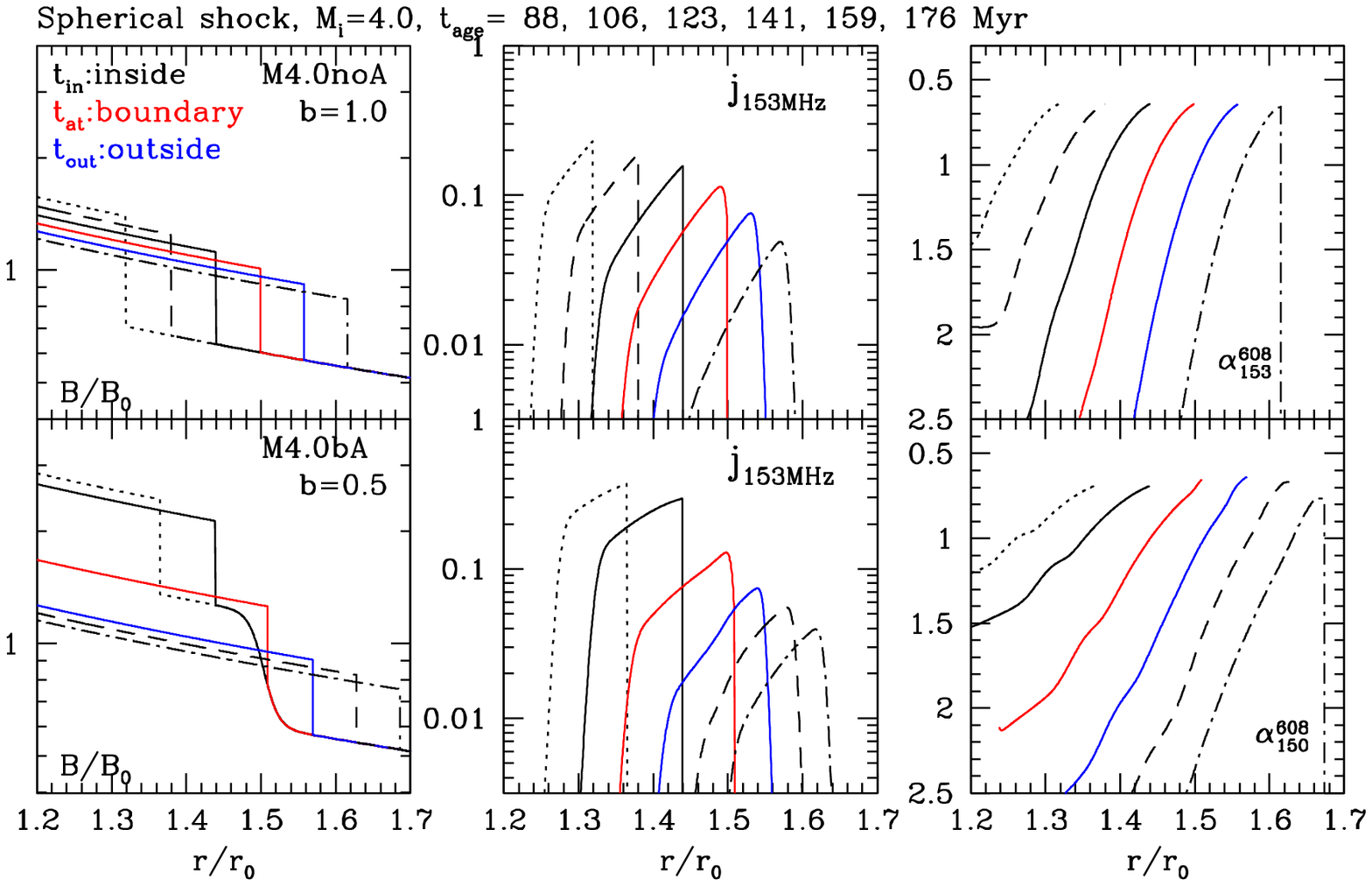}
\caption{Time evolution of $M_{\rm s,i}=4.0$ shocks without ($b=1.0$, upper panels) and with ($b=0.5$, lower panels) a bubble.
Magnetic field strength (left panels), synchrotron emissivity at 150~MHz (middle panels), 
spectral index, $\alpha_{153}^{608}$ between 153 and 608~MHz (right panels) are shown at six different time epochs,
{ $t_{\rm age} =$ 88, 106, 123, 141, 159, and 179~Myr (the shock expands radially outward).
Here $r$ is the radial distance from the cluster center.
The shock is located inside the bubble at $t_{\rm in}$ (black solid lines), at the boundary zone at $t_{\rm at}$ (red solid lines),
and outside the bubble at $t_{\rm out}$ (blue solid lines) in both models.
\label{fig5}}
}
\end{figure*}
%---------------------------------------------------------------

Observed polarization vectors of radio emission indicate that the overall magnetic field directions behind the Sausage and Toothbrush relics are roughly quasi-perpendicular, i.e., parallel to the shock surface \citep{vanweeren10, vanweeren12}.
This seems consistent with the findings of plasma simulations by \citet{guo2014} that electrons may be injected and accelerated more efficiently at quasi-perpendicular shocks.
Here we assume that the preshock magnetic fields in the ICM or inside the bubble are mostly turbulent and isotropic, and they become quasi-perpendicular downstream through compression of the transverse components across the model shock:
\begin{equation}
B_2=B_1 \sqrt{1/3+2\sigma^2/3},
\label{B2B1}
\end{equation}
where $B_1$ and $B_2$ are the immediate preshock and postshock magnetic field strengths, respectively.
{ Using MHD simulations, \citet{ji2016} studied magnetic field amplification through turbulent dynamo at shocks with the parameters relevant for ICM shocks.
In the case of low $M_{\rm s}$ shocks, they found that amplification is dominated mostly by compression, so the magnetic field model in Equation (\ref{B2B1})
could be valid for an isotropic tangled preshock field.

As in our previous studies \citep{kang16a, kang16b, kang17}, we further assume} that the magnetic energy is dissipated behind the shock, so the downstream field strength decreases with the gas pressure away from the shock as
\begin{equation}
B_{\rm dn}= B_2 [P(r)/P_2]^{1/2}.
\label{Bdn}
\end{equation}
In addition, we also consider an alternative model where the postshock magnetic field decreases radially similarly to $B_{\rm ICM}$,
\i.e., $B_{\rm dn}(r)= B_2 \cdot (r/r_s)^{-1.25}$, where $r_s$ is the shock radius.

In summary, we consider the following three models for the magnetic field profiles (see Figure 3):\\
A: $B_{\rm ICM}=B_0 (r/r_0)^{-1.25}$, $B_{b}\propto n_{\rm b}^{-1} $, $B_{\rm dn} \propto r^{-1.25}$,\\
B: $B_{\rm ICM}=B_0$, $B_{\rm b}\propto n_{\rm CRe}$, $B_{\rm dn}\propto P^{1/2}$,\\
C: $B_{\rm ICM}=B_0$, $B_{\rm b}=B_0$, $B_{\rm dn}\propto P^{1/2}$.\\
{ Here $B_0 = 1 \muG$ is the reference field strength at the initial shock location ($r_s/r_0=1.0$) as shown in Figure~\ref{fig3}.
The mode C is the magnetic field configuration adopted in \citet{kang16a, kang16b, kang17}.}

For a given shock compression ratio $\sigma$, $B_2$ is controlled by $B_1$, which is in turn specified by the bubble field $B_{\rm b}(r)$ at the shock position, $r_s(t)$, when the shock travels inside the bubble.
Then the surface brightness profile, $I_{\nu}(R)$ behind the shock, depends mainly on $B_{\rm dn}$ that governs the radiative cooling rates of postshock electrons and synchrotron radiation flux.
{ Here $R$ is the distance behind the projected shock edge in the plane of the sky as defined in Figure~1 of \citet{kang15}.
For simplicity, we assume that the electron pitch angle distribution is isotropic with respect to local magnetic field directions, when we estimate synchrotron cooling and emission.
}

\section{Results of DSA Simulations\label{results}}
%---------------------------------------------------------------
\begin{figure*}[t!]

\centering
\includegraphics[trim=2mm 40mm 6mm 5mm, clip, width=160mm]{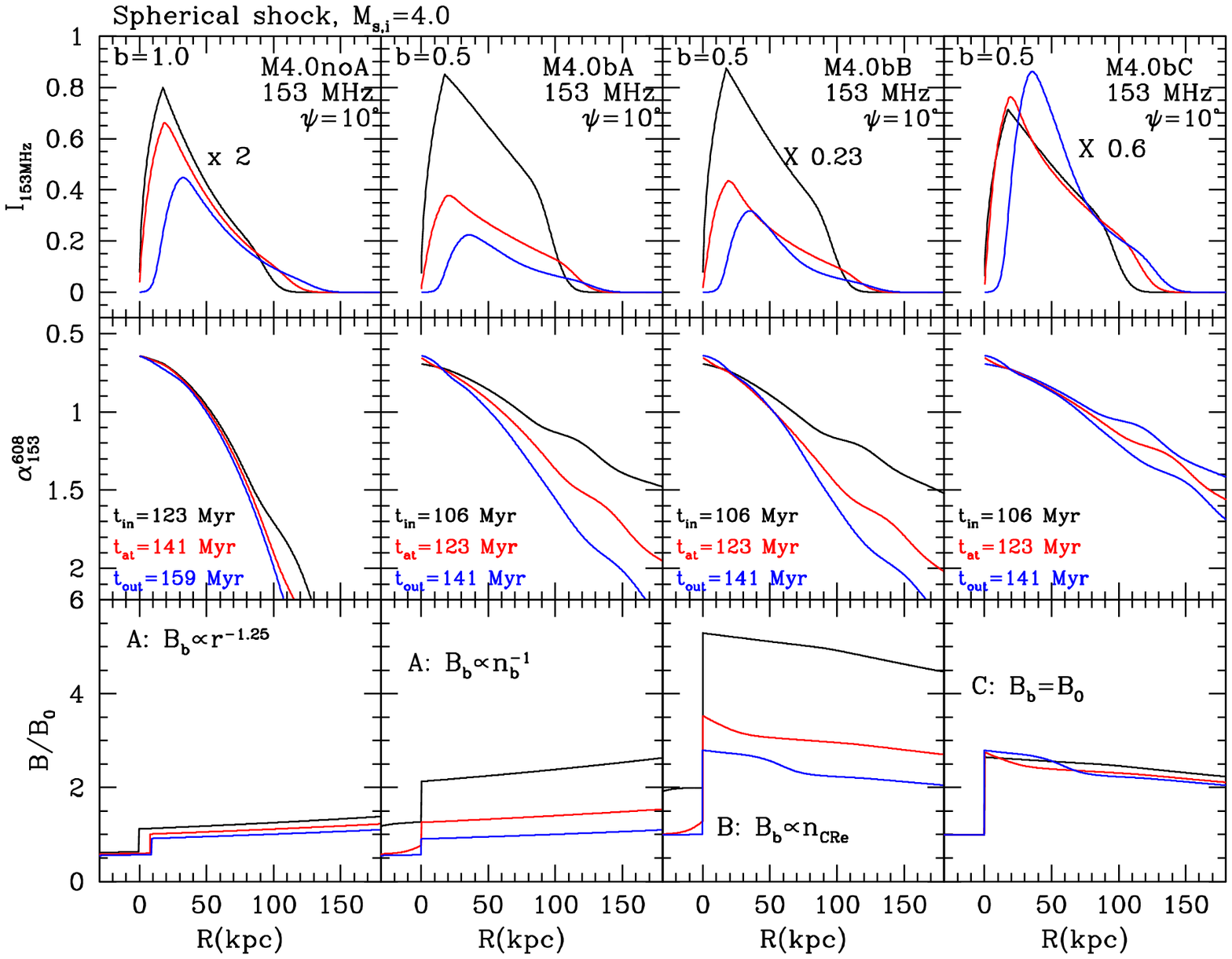}
\caption{Postshock profiles of the surface brightness, $I_{\nu}(R)$ at 153~MHz (top panels) in arbitrary units,
the spectral index $\alpha_{153}^{608}(R)$ between 153~MHz and 608~MHz (middle panels),
and the magnetic field (bottom panels) for the M4.0noA, M4.0bA, M4.0bB, and M4.0bC models.
For some models, $I_{\nu}$'s are scaled by the numerical factors specified in the figure.
{ Here $R$ is the distance in units of kpc behind the projected shock edge ($R=0$) in the plane of the sky.}
The extension angle, $\psi=10^{\circ}$, is adopted.
{ As in Figure~\ref{fig5}, the shock is located inside the bubble at $t_{\rm in}$ (black), in the boundary zone at $t_{\rm at}$ (red),
and outside the bubble at $t_{\rm out}$ (blue) in all the models.}
Three shock ages are $t_{\rm in,at,out}= 123,~141,~159$~Myr for M4.0noA and $t_{\rm in,at,out}= 106,~123,~141$~Myr for M4.0bA, M4.0bB, and M4.0bC.
The initial profile of $B(r)$ for each model is shown in the right panel of Figure~\ref{fig3}.\label{fig6}
}
\end{figure*}
%---------------------------------------------------------------

Table~\ref{table1} shows the parameters for model shocks with initial Mach number $M_{\rm s,i}=4.0$ considered in this study.
In the M4.0noA model, the density parameter $b=1.0$ means that the cloud containing fossil electrons have the same density and temperature as the surrounding ICM.
Magnetic field model A is assumed, so $B(r)\propto r^{-1.25}$ with a jump at the shock location (see Figure~\ref{fig5}).
The M4.0bA, M4.0bB, and M4.0bC models with $b=0.5$ have the magnetic field profiles A, B, and C, respectively, as shown in Figure \ref{fig3}.

Figure~\ref{fig5} shows the evolution of the magnetic field, synchrotron emissivity at 153~MHz, and spectral index, $\alpha_{153}^{608}$ between 153 and 608~MHz
for the M4.noA ($b=1.0$) and M4.0bA models ($b=0.5$) { at six different shock ages specified in the figure.
Among them, we choose three epochs that correspond to times when the shock is inside ($t_{\rm in}$), at the boundary ($t_{\rm at}$), and outside ($t_{\rm out}$) of the bubble in order to compare the two models. Due to the higher shock speed inside the bubble, these three epochs are different in the two models, that is, $t_{\rm in,at,out}= 123,~141,~158$~Myr for the M4.0noA model, while
$t_{\rm in,at,out}= 106,~123,~141$~Myr for the M4.0bA model.}

The downstream magnetic field profile is assumed to be $B_{\rm dn} = B_2 (r/r_s)^{-1.25}$ in both models,
while $B_2$ is given by Equation (\ref{B2B1}).
For the bubble field, $B_{\rm b}$, the model A is adopted, so the preshock field, $B_1$ and the downstream field, $B_{\rm dn}$, are stronger during the time when the shock is inside the bubble in the M4.0bA model, compared to the ICM field in the M4.0noA model.

In the M4.0bA model, when the shock inside the bubble ($t_{\rm age} \le t_{\rm in}$), $j_{153}$ is higher due to stronger $B_{\rm dn}$, and $\alpha_{153}^{608}$ is
slightly larger due to smaller $M_{\rm s,b}$, compared to those in the M4.0noA model.
As the shock exits the bubble ($t_{\rm age} \ge t_{\rm at}$), both $B_1$ and $B_{\rm dn}$ decrease in time,
so the synchrotron emissivity $j_{153}$ drops significantly.
Although $B_{\rm dn}$ is higher in the M4.0bA model, the downstream widths of $j_{153}$ and $\alpha_{153}^{608}$ are wider than those in the M4.0noA model,
because the shock speed is higher inside the hotter bubble.

%---------------------------------------------------------------
\begin{figure*}[t!]
\centering
\includegraphics[trim=2mm 120mm 5mm 5mm, clip, width=160mm]{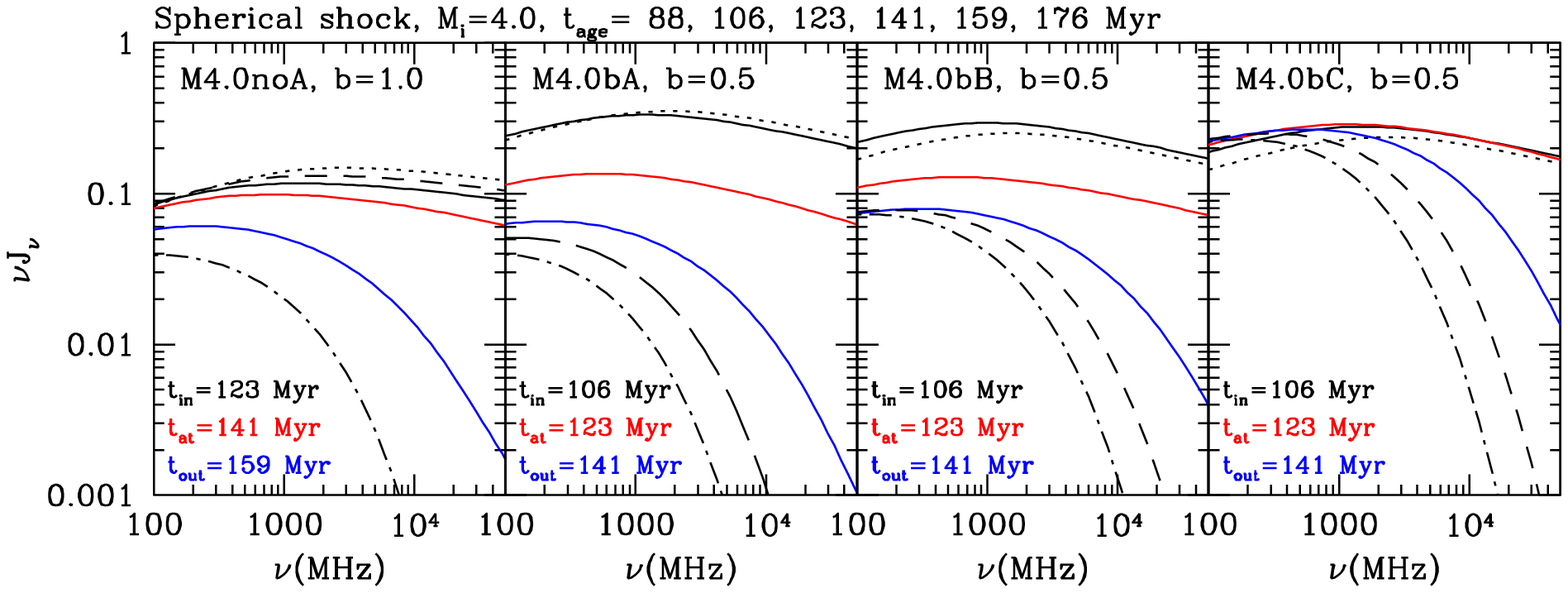}
\caption{Time evolution of volume-integrated radio spectrum for M4.0noA, M4.0Ba, M4.0Bb, and M4.0Bc models
are shown at six different time epochs, {$t_{\rm age} =$ 88, 106, 123, 141, 159, and 179 Myr as in Figure~\ref{fig5}.}
The three shock ages, $t_{\rm in,at,out}$, are the same as in Figure~\ref{fig6}.
\label{fig7}
}
\end{figure*}
%---------------------------------------------------------------

\subsection{Surface Brightness and Spectral Index Profiles\label{brightness}}

Using the CRe energy spectrum and the magnetic field strength in the DSA simulations,
we first calculate the synchrotron emissivity $j_{\nu}(r)$ of each spherical shell.
The radio surface brightness, $I_{\nu}(R)$, is calculated by integrating $j_{\nu}(r)$ along a given line-of-sight,
where a postshock volume of radio-emitting electrons, shaped like a spherical wedge, is adopted, as in Figure 1 of \citet{kang15}.
This volume is specified by the extension angle, $\psi=10^{\circ}$.
{ Again note that $R$ is the distance behind the projected shock edge in the plane of the sky.}

The top panels of Figure~\ref{fig6} show $I_{\rm 153MHz}(R)$ (in arbitrary units) at 153~MHz for the M4.0noA, M4.0bA, M4.0bB, and M4.0bC models,
while the middle panels show the spectral index profile, $\alpha_{153}^{608}(R)$, calculated with the ratio between
$I_{\rm 153MHz}(R)$ and $I_{\rm 608MHz}(R)$.
In this figure the shock faces to the left, {so the region of $R>0$ is the postshock region that could be observed as a radio relic across the plane of the sky.}
Because magnetic field strengths are different for different models as shown in the bottom panels,
the scale of $I_{\rm 153MHz}$ is adjusted by the numerical factors given
in the top panels, but the flux levels of the three curves in each panel reflect their relative strengths.

For each model, the results are shown at $t_{\rm in,at,out}$ as the shock exits out of the bubble.
For $t_{\rm age}\le t_{\rm in}$ the shock position ($R=0$) corresponds to the relic edge
and the results are similar to those at $t_{\rm in}$.
For $t_{\rm age}\ge t_{\rm out}$, on the other hand,
because the shock sweeps through the ICM with $n_{\rm CRe}=0$ and fossil electrons are no longer injected at the shock,
the relic edge lies further downstream behind the shock (to the right in the figure).
The spectral index at the shock position, $\alpha_{\rm sh}$, decreases slightly from $t_{\rm in}$ (black) to $t_{\rm out}$ (blue),
because the shock Mach number increases as the shock exits out of the bubble (see the lower right panel of Figure~\ref{fig4}).

For shock speeds, $u_s\approx 2.7\times 10^3 \kms$ and $u_2 \approx 8.0\times 10^2 \kms$,
the advection length behind the shock $l_{\rm adv}=u_2 t_{\rm age} \approx 85$, 100, and 110 ~kpc at $t_{\rm age}\approx 106$, 123, and 141~Myr, respectively,
which is consistent with the black, red, and blue lines for the M4.0bA model in Figure~\ref{fig6}.
Both advection and cooling lengths scale with postshock flow speed, so models with bubbles have longer widths behind the edge, compared to the M4.0noA model.
The characteristic energy of electrons emitting at the peak frequency, $\nu_{\rm peak} =153$~MHz, in $B_2\approx 3 \muG$ is $\gamma_e \approx 6.4 \times 10^3$,
and the cooling time scale of those electrons is $t_{\rm rad}\approx 100$~Myr.
Since the shock ages shown in the figure are similar to or slightly larger than $t_{\rm rad}$, radiative cooling becomes important for $t_{\rm age} \gtrsim t_{\rm in}$.
The cooling length scale for $\gamma_e \approx 6.4 \times 10^3$ electrons is $l_{\rm cool}=u_2 t_{\rm rad}\sim 110$~kpc \citep[See Eq. (6) in][]{kang16a}.

As shown in the bottom panels, in the M4.0bA and M4.0bB models the preshock and downstream magnetic fields decrease in time,
as the shock runs through the bubble boundary from $t_{\rm in}$ to $t_{\rm out}$. So $I_{\rm 153MHz}(R)$ declines
and $\alpha_{153}^{608}(R)$ steepens rapidly during that period.
In the M4.0bC model, on the other hand, $B_1$ and the ensuing downstream field $B_{\rm dn}$ remain almost constant in time,
so $I_{\rm 153MHz}$ remains high and the profiles of $\alpha_{153}^{608}(R)$ do not change significantly for $t_{\rm age}>t_{\rm out}$.
{ The figure demonstrates that obviously the amplitude and profile of $I_{\nu}(R)$
depends on the detailed models for $n_{\rm CRe}(r)$, $B_{\rm dn}(r)$, and the path length along the line of sight (i.e., $\psi$),
as well as shock dynamics.}

\subsection{Integrated Spectrum\label{spectrum}}

Figure~\ref{fig7} shows the time evolution of the volume-integrated spectrum, $\nu J_{\nu}$, for the M4.0noA, M4.0bA, M4.0bB, and M4.0bC models.
Again, the black, red, and blue lines show $J_{\nu}$ at $t_{\rm in}$, $t_{\rm at}$, and $t_{\rm out}$, respectively, as shown in Figure~\ref{fig5}.
The integrated spectrum depends on both $n_{\rm CRe}$ and $B_{\rm dn}$ as well as shock dynamics.
In the case of the M4.0bC model the evolution of $J_{\nu}$ can be explained simply by disappearance of fossil electrons outside the bubble,
because $B_{\rm dn}$ is almost constant in time.
When the shock is still inside the bubble for $t_{\rm age} \le t_{\rm in}$ (e.g., two black lines in the top),
the integrated spectrum steepens approximately from $\nu^{-\alpha_{\rm sh}}$ to $\nu^{-(\alpha_{\rm sh}+0.5)}$ at $\sim$GHz.
As the shock exits the bubble from $t_{\rm at}$ (red) to $t_{\rm out}$ (blue), $n_{\rm CRc}$ decreases to zero,
so $J_{\nu}$ declines in amplitude and steepens above GHz with the break frequency getting progressively lower in time.
In the M4.0noA model with $b=1.0$, the preshock field $B_1=B_{\rm ICM}(r_s)$ decreases slightly outward,
so $J_{\nu}$ decreases a bit more, compared to M4.0bC model, due to
the combined effects of lack of fossil electrons and small decline in magnetic fields.

In the M4.0bA and M4.0bB models, on the other hand, the significant reduction of both $n_{\rm CRc}$ and $B_{\rm dn}$
leads to rapid decline of $J_{\nu}$ from $t_{\rm in}$ to $t_{\rm out}$, as the shock exits out of the bubble.
This exercise demonstrates that it is important to understand how fossil electrons are deposited in the ICM
and how the downstream magnetic field decays or amplifies behind the shock
in understanding radio observations of radio relics.

\section{Summary\label{summary}}

Radio jets from AGNs can deposit relativistic electrons in the ICM of galaxy clusters.
These jets further expand into the ICM, producing lobes/plumes on $\sim 100$~kpc scales,
which can be transported radially to the cluster outskirts by the buoyant force.
Through entrainment of the surrounding medium, they are expected to become gas bubbles of lower-density but higher-temperature, relative to the ICM,
and may contain dynamically-insignificant populations of fossil relativistic electrons.
In this study, we consider a feasible model in which a radio relic is generated when a merger-driven shock with $M_{\rm s} \approx 4$ sweeps through such a bubble with seed CR electrons to be injected to the DSA process.
Shock-accelerated electrons radiate diffuse synchrotron emission in $\sim \muG$ magnetic fields.
This scenario is consistent with the observational fact that
only a small fraction ($\sim 10\%$) of merging clusters host radio relics \citep{feretti12}.

{Inevitably, introduction of a hotter bubble with $\rho_{\rm b}$, $n_{\rm CRe}$ and $B_{\rm b}$ increases the number of free parameters in modeling of radio relics, which could lead to much richer dynamics in such models.
With this caveat we summarize the main findings of this study as follows}:

1. For a bubble with moderate density parameter,  $b=n_{b}/n_{\rm ICM}=T_{\rm ICM}/T_{b}\sim 0.5$, the shock speed increases by 20\% but
the shock Mach number decreases by 15\% as the shock runs into the hotter bubble.
As a result, the width of the radio-emitting region (radio relic) behind the shock is larger due to the larger postshock flow speed,
but the radio spectral index at the relic edge, $\alpha_{\rm sh}$, is slightly larger due to the lower shock Mach number.
As the shock exits the bubble, on the other hand, the shock slows down roughly to the original speed, while the shock Mach number increases to the value similar to that before the shock meets the bubble.

2. The surface brightness, $I_{\nu}$, and spectral index, $\alpha_{\nu}$, of radio relics depend sensitively on
{ the spatial distributions  of downstream magnetic field, $B_{\rm dn}(r)$, and fossil electron density, $n_{\rm CR}(r)$,}
in addition to the shock speed and Mach number (see Figure~\ref{fig6}).
So it is crucial to understand how $B_{\rm dn}$ is amplified or decay via various MHD/plasma processes behind the shock, which is beyond the scope of this study.
Here, instead, we consider three different models for the magnetic field profiles as shown in the right panel of Figure~\ref{fig3}.
As in most of the previous studies, in the M4.0bC model we assume that the preshock magnetic field, $B_1$, is constant and the downstream field, $B_{\rm dn}\propto \sqrt{P}$, is almost constant, so the spectral steepening profile behind the shock remains almost the same in time.

3. The magnetic field strength in the cluster outskirts is observed to decrease with the gas density, for example, $B_{\rm ICM} \propto n^{0.5} \propto r^{-1.25}$.
Hence, it is rational to expect that the preshock magnetic field strength declines during the formation of radio relics,
considering that the shock with $M_{\rm s}\approx 3$ must sweep at least $\sim 300-450$~kpc of the region with fossil electrons in order to explain
typical widths of giant radio relics, $\sim 100-150$~kpc.
In addition, bubbles generated by radio galaxies could have stronger magnetic fields than the surrounding ICM.
So more realistic pictures would be the magnetic field models in the M4.0bA or M4.0bB models, in which the preshock field $B_1$ and the downstream $B_{\rm dn}$ change as the shock propagates through the bubble.

4. The spectral curvature of the volume-integrated spectrum of radio relics is governed not only by the magnetic field profile behind the shock but also by the {spatial distribution} of fossil electrons, in addition to shock dynamics.
Thus it is necessary to understand those physical properties in the bubble and the ICM in order to establish a realistic picture for the origin of radio relics in galaxy clusters.

%--------------------------------------------------------------------
\acknowledgments{
This work was supported by a 2-Year Research Grant of Pusan National University.
The author thanks D. Ryu for helpful discussions on shock dynamics in low-density bubbles.

%--------------------------------------------------------------------

%-------------------------------------------------------------------
\end{document}